\documentclass[twoside,12pt]{article}

\usepackage{amsmath,amssymb,endnotes}

\newcounter{muni}
\newenvironment{remunerate}{\begin{list}{{\rm \arabic{muni}.}}
{\usecounter{muni}\setlength{\leftmargin}{0pt}
\setlength{\itemindent}{38pt}}}{\end{list}}

\newcommand{\nc}{\newcommand}                  \nc{\nf}{\infty}    
\nc{\dst}{\displaystyle}                       \nc{\nnb}{\nonumber} 
\nc{\beq}{\begin{equation}}                    \nc{\eeq}{\end{equation}} 
\nc{\beqa}{\begin{eqnarray}}                   \nc{\eeqa}{\end{eqnarray}}
\nc{\brm}{\begin{remunerate}}                  \nc{\erm}{\end{remunerate}}
\nc{\barr}{\begin{array}}                      \nc{\earr}{\end{array}}
             \nc{\mc}{\mathcal}

\nc{\bs}{\backslash}        \nc{\nl}{\newline}      \nc{\mb}{\mathbb}
\nc{\qq}{\quad\quad}        \nc{\ol}{\overline}     \nc{\pt}{\partial}
\nc{\dg}{\dagger}

\nc{\alf}{\alpha}    \nc{\be}{\beta}        \nc{\ga}{\gamma}   \nc{\si}{\sigma} 
\nc{\de}{\delta}     \nc{\eps}{\epsilon}    \nc{\vtht}{\vartheta}
\nc{\om}{\omega}     \nc{\vp}{\varphi}    \nc{\vsi}{\varsigma}
\nc{\vrho}{\varrho}  \nc{\tht}{\theta}      \nc{\la}{\lambda}

\nc{\Om}{\Omega}     \nc{\Ga}{\Gamma}      \nc{\De}{\Delta}

\nc{\Log}{{\rm Log }}           \nc{\tg}{{\rm tg }}
\nc{\sh}{{\rm sh }}             \nc{\ch}{{\rm ch }}

\nc{\tr}{{\rm tr }}

\nc{\LIM}{\mathop{\smash{\rm LIM}}}

\title{\bf Renormalisability of  non-homogeneous T-dualised sigma-models}
\author{Pierre-Yves Casteill \thanks {\noindent
Laboratoire de Physique Th\'eorique et des Hautes Energies,
 Unit\'e associ\'ee au CNRS UMR 7589, Universit\'e Paris 7,
 2 Place Jussieu, 75251 Paris Cedex 05. casteill@lpthe.jussieu.fr}}

\begin{document}
\date{\today}
\maketitle

\begin{abstract}
The quantum equivalence between $\sigma$-models and their non-abelian T-dualised partners is 
examined for a large class of four dimensional non-homogeneous and quasi-Einstein metrics with 
an isometry group $SU(2)\times U(1)$. We prove that the one-loop renormalisability of the 
initial torsionless $\sigma$-models is equivalent to the one-loop renormalisability of 
the T-dualised torsionful model. For a subclass of K\"ahler original metrics, the 
dual partners are still K\"ahler (with torsion).
\end{abstract}



\section{Introduction}
The subject of target space duality, or T-duality, in String Theory and in Conformal Field Theory has generated much interest in recent years and extensive reviews covering abelian, non-abelian dualities and their applications to string theory and statistical physics are available in the literature \cite{aab1,aal3,gpr}. The geometrical aspects of this duality can be found in \cite{Al1}. T-duality provides a method for relating inequivalent string theories. First discovered for the case of $\sigma$-models with some abelian isometry, the concept of T-duality has been recently enlarged to theories with non-abelian isometries \cite{aal2,Lo,oq}. A very important and interesting property of T-duality applied on non-abelian isometry is that it can map a geometry with such isometries to another which has none. Therefore, non-abelian T-duality can not be inverted as in the abelian case. 

By showing that T-duality is a canonical transformation \cite{CZ,grav,aal2}, it was proved that theories in such way related where classically equivalent. Furthermore, this equivalence was still remaining at the one-loop level, in a strict renormalisability sense, in all the many example that have been tested up to now to this duality, with an emphasis put on $SU(2)$ \cite{fj,ft,aab1,aal1,gr,oq}. For example, this one-loop equivalence still remains for principal $\sigma$-models whatever strongly broken the right isometries may be \cite{CV}. The non-abelian dualisation of non-homogeneous metrics such as the Schwarzschild black hole or Taub-NUT was performed in \cite{oq},\cite{aal1} and in \cite{He}. We propose here the dualisation of the general $SU(2)\times U(1)$ metrics.

Problems arise when one addresses the question of the renormalisability of dualised theories beyond the one-loop order. It had been proved that even for the simplest $(SU(2)\times SU(2))/SU(2)$ principal $\sigma$-model, the dualised theory is not two-loop renormalisable, in the minimal dimensional scheme \cite{st,kt}. However, as shown in \cite{BC}, a \emph{finite} deformation at the $\hbar$ order of the dualised metric is sufficient for recovering a two-loop renormalisability for this particular model. As it will be shown, the $SU(2)\times U(1)$ $\sigma-$models are not in general two-loop renormalisable, even though the one-loop renormalisability remains for their dual partners !

The content of this article is the following : in section 2, we recall the general expression of the $SU(2)\times U(1)$ metrics and set the notations. In section 3, we make a review of such metrics which give rise to one-loop renormalisable $\sigma-$models, as for example the celebrated Taub-NUT and Eguchi-Hanson metrics. In section 4, we show that only the particular metrics where homogeneity is recovered by some enhancement of the isometries are two-loop renormalisable. In section 5, we dualise the original theory and show in section 6 that the one-loop renormalisability survives during the dualisation process. When the original metric is K\"ahler, we investigate in section 7 if such a property is still present for the dual partner. Some concluding remarks are offered in section 8.

\mathversion{bold}
\section{The $SU(2)\times U(1)$ metric}
\mathversion{normal}

We consider the four dimensions metrics with cohomogeneity one under a $SU(2)\times U(1)$ isometry. In the more general way, these can write 
\begin{equation*}
g =\alpha(t)\,dt^2+\beta(t)\,({\sigma_1}^2+{\sigma_2}^2)+\gamma(t)\,{\sigma_3}^2 \ ,
\end{equation*}
where the $\sigma_i$ are 1-forms such that
\begin{equation*}
d\sigma_i=\varepsilon\,\frac12\,\epsilon_{ijk}\,\sigma_j\wedge\sigma_j\ ,\qq\varepsilon=\pm1\ .
\end{equation*}
One can always writes ${\sigma_1}^2+{\sigma_2}^2$ and $\sigma_3$ under the well known specific shape
$${\sigma_1}^2+{\sigma_2}^2=d\theta^2+\sin^2\theta\,d\varphi^2\ ,\qq\sigma_3=d\psi+\cos\theta\,d\varphi\ .$$
If $\varepsilon=+1$, the triplet of 1-forms $\vec \sigma$ is changed under infinitesimal transformations of $su(2)_L\oplus~su(2)_R$ as
$$\delta \vec\sigma=\vec\epsilon_R\wedge\vec\sigma\ .$$
Therefore $\vec\sigma$ is a $SU(2)_L$ singlet and a $SU(2)_R$ triplet. If $\beta(t)\neq\gamma(t)$, the $SU(2)_R$ isometries will be broken down to a $U(1)$ and the total isometry group of the metric will then be $SU(2)_L\times U(1)$. Indeed, in order to keep the metric invariant, one then must have $\vec \epsilon_R=\{0,0,\mu\}$.
If $\varepsilon=-1$, $\vec \sigma$ is changed under infinitesimal transformations of $su(2)_L\oplus~su(2)_R$ as
$$\delta \vec\sigma=\vec\epsilon_L\wedge\vec\sigma\ ,$$
and therefore the isometry group of the metric will be $SU(2)_R\times U(1)$.
 The choice of $\varepsilon$ switches also the autodual components of the Weyl tensor ($W_+\leftrightarrow~ W_-$). In all cases, when $\beta(t)=\gamma(t)$, the metric has for isometry group $SU(2)_L\times SU(2)_R$ and is conformally flat.

It is then possible to define the $\sigma$-model corresponding to these metrics
\begin{equation}\label{actioni}
S=\frac{1}{T}\int dx^2\,\eta^{\mu\nu}\, g_{ij}\,\partial_\mu\phi^i\,\partial_\nu\phi^j\ ,
\end{equation}
with $\{\phi^0=t,\phi^1=\theta,\phi^2=\varphi,\phi^3=\psi\}$, and address the question of its one-loop and two-loop renormalisability.

In order to derive the Ricci tensor, we define the vierbein $\{e_a|a\in\{0,1,2,3\}\}$ as
$$\begin{array}{rlrl}
 e_0 &=\sqrt{\alpha(t)}\,dt \ ,&\qq\qq e_2 &=\sqrt{\beta(t)}\,\sigma_2 \ ,\nnb\\
e_1 &=\sqrt{\beta(t)}\,\sigma_1 \ ,&\qq\qq e_3 &=\sqrt{\gamma(t)}\,\sigma_3\ .
\end{array}$$

In the absence of torsion, the condition for giving one-loop renormalisability is the quasi-Einstein property of the metric :
\begin{equation}
\label{qe}
Ric_{ab}=\lambda\,g_{ab}+D_{(a}v_{b)}\ ,
\end{equation}
where the Einstein constant $\lambda$ will renormalise the coupling while the vector  $v$ will renormalise the field. 

\section{One-loop renormalisation}

We will only consider metrics satisfying condition (\ref{qe}) so that the corresponding $\sigma$-models are one-loop renormalisable. Of course, as we want to keep the $SU(2)$ symmetry while renormalising, we will only consider here vectors $v$ that depends only on the $t$ coordinate : $v=v(t)$.
As the expression of the $SU(2)\times U(1)$ metric (\ref{meti}) we chose does not mix $dt$, $\sigma_1$, $\sigma_2$ and $\sigma_3$, both the metric $g$ and the Ricci tensor $Ric$ will be diagonal in the $\{dt,\sigma_1,\sigma_2,\sigma_3\}$ basis and this will hold in the vierbein. As a consequence, $D_{(a}v_{b)}$ must be also diagonal ; this is true only for vectors of the form $v=v_0(t)\,e_0+\rho\,\sqrt{\gamma(t)}\,e_3\,$. The constant $\rho$ is arbitrary as $\sqrt{\gamma(t)}\,e_3$ is in fact the form dual to the Killing $\pt_\psi$. We will take $\rho=0$.

In order to simplify matters, from now on, we will choose the  coordinate $t$ so that  $\beta(t)=t$. The metric now writes
\begin{equation}
\label{meti}
g=\alpha(t)\,dt^2+t\,({\sigma_1}^2+{\sigma_2}^2)+\gamma(t)\,{\sigma_3}^2\ .
\end{equation}

All this being settled, the quasi-Einstein character of the metric (\ref{qe}) can now be expressed as a set of three non-linear differential equations which are :

\begin{equation} \label{eq}
\left\{
\begin{array}{rll}
\dst \frac{1}{t^2}  + \left(\frac{1}{t}+\frac{\gamma '(t)}{2\,\gamma (t)} \right)\, \frac{\alpha '(t)}{\alpha (t)} + \frac{{\gamma '(t)}^2}{2\,{\gamma (t)}^2} - \frac{\gamma ''(t)}{\gamma (t)}
&=& \dst 2\,\lambda \,\alpha (t) + 2\,{\sqrt{\alpha (t)}}\,v_0'(t)
\\[8mm]
\dst  2\,\left(2\,  - 
  \frac{\gamma (t)}{t}\right)\,\alpha (t) + \frac{\alpha '(t)}{\alpha (t)} - \frac{\gamma '(t)}{\gamma (t)}
 &=& \dst 4\,\lambda\,t\,\alpha (t) + 2\,{\sqrt{\alpha (t)}} \,v_0(t)
 \\[8mm]
\dst -\frac{2}{t} + 
 \frac{2}{t^2}\,\frac{{\gamma (t)}^2}{\gamma '(t)}\,\alpha (t) + \frac{\alpha '(t)}{\alpha (t)} + \frac{\gamma '(t)}{\gamma (t)} - 
  \frac{2\,\gamma ''(t)}{\gamma '(t)}  
&=& \dst 4\,\lambda \,\frac{\gamma (t)}{\gamma '(t)}\,\alpha (t) + 2\,{\sqrt{\alpha (t)}}\,v_0(t)
\end{array}
\right.
\end{equation}

This system is difficult to solve, even though it can still be done for some limited cases as the Einstein one ($v_0=0$) and the quasi-Einstein K\"ahler one. It is possible to eliminate $\alpha(t)$ and $v_0(t)$ in the system (\ref{eq}), leading to a single, deeply non-linear, differential equation of the fourth order in $\gamma(t)$. The general $SU(2)\times U(1)$ quasi-Einstein metric should therefore depend on four parameters.

In order to convince the reader of the large class of models that will be dualised, we will now give a short review of the $SU(2)\times U(1)$  Einstein and quasi-Einstein K\"ahler metrics.

\vspace{0.3cm}
\begin{flushleft}
\textbf{Einstein metrics :}
\end{flushleft}

The metric $g$ will be Einstein if  $Ric=\lambda\,g$. It is possible to integrate the differential system (\ref{eq}) imposing $v_0=0$ and one gets
\begin{equation}\label{original}
\alpha(t) = \frac{1}{1+A\,t}\cdot\frac{1}{\gamma(t)}\ ,\qq
\gamma(t) = \frac{4\,t}{{\left( 1 + {\sqrt{1 + A\,t}} \right) }^2} -  \frac{4\,\lambda\,t^2}{3}\,
  \frac{3 + {\sqrt{1 + A\,t}}}{{\left( 1 + {\sqrt{1 + A\,t}} \right) }^3} + \frac{B}{t}\,{\sqrt{1 + A\,t}} \ ,
\end{equation}
$A$ and $B$ being the integration constants. This family contains many metrics of interest which we recall briefly.

If $A=0$, we recover the K\"ahler-Einstein extension of Eguchi-Hanson \cite{gp}.
If $A \neq 0$ then $g$ identifies with the large class of Einstein metrics derived by Carter \cite{carter}. By making the change of coordinates,
\begin{equation*}
t \longrightarrow t^2-n^2\ ,\qq\hbox{with}\ 
A =\frac1{n^2}\hbox{ and }
B =-8\,(M-n)\,n^3\ ,
\end{equation*}
one can have for $g$ a more simple expression~:
\begin{equation}\label{TN}
\left\{
\begin{array}{rll}
g &=&\dst \frac{t^2-n^2}{f(t)}\,dt^2+(t^2-n^2)\,({\sigma_1}^2+{\sigma_2}^2)+\frac{4\,n^2}{t^2-n^2}\,f(t)\,{\sigma_3}^2\ ,\\[8mm]
f(t) &=&\dst t^2-2\,M\,t+n^2-\frac{\lambda}{3}\,(t-n)^3(t+3\,n)\ .
\end{array}
\right.
\end{equation}
Notice that as $A$ and $B$ are real constants, $M$ and $n$ can be both reals or pure imaginaries.
Defining $2\,n\,d\psi=d\Psi$ and taking the limit $n\rightarrow0$ gives the Schwarzschild metric with cosmological constant :
\begin{equation}\label{schwar}
g =\frac{1}{\dst1-\frac{2\,M}{t}-\frac{\lambda}{3}\,t^2}\,dt^2+t^2\,(d\theta^2+\sin^2\theta\,d\varphi^2)+\left(1-\frac{2\,M}{t}-\frac{\lambda}{3}\,t^2\right)\,d\Psi^2\ .
\end{equation}

Other limits of (\ref{TN}) lead to the Page metric on $P_2(\mathbb{C})\#\overline{P_2(\mathbb{C})}$ and to the Taub-NUT metric.

\begin{flushleft}
\textbf{Quasi-Einstein K\"ahler metrics :}
\end{flushleft}

These are the only  $SU(2)\times U(1)$ quasi-Einstein metrics known up to now \cite{chave}. We suppose here that there is a choice of holomorphic coordinates on which the isometries $SU(2)\times U(1)$ act linearly. It happens that this hypothesis implies the integrability of the complex structure.
A necessary condition of the K\"ahler property is the closing of the K\"ahler form :
$$d(e_0\wedge e_3 +\varepsilon\,e_1\wedge e_2)=d\left(\sqrt{\alpha(t)\,\gamma(t)}\,dt\wedge\sigma_3+\beta(t)\,d\sigma_3\right)=0\ .$$
It is clear that this relation will hold iff ${\beta'(t)}^2=\alpha(t)\,\gamma(t)$, {\it i.e.} $\dst \alpha(t)=\frac{1}{\gamma(t)}$. It is then possible to solve system (\ref{eq}) and one gets for the metric and for the vector $v$ :
\begin{equation}\label{ka}
g=\frac{1}{\gamma(t)}\,dt^2+t\,\left({\sigma_1}^2+{\sigma_2}^2\right)+\gamma(t)\,{\sigma_3}^2\,,\qq v=-C\,\sqrt{\gamma(t)}\,e_0=-C\,dt\,,
\end{equation}
with
$$\gamma(t)=\frac{D\,e^{C\,t}}{t} + t + \frac{2 }{C^2\,t} \,
     \left( 1 - \frac{2\,\lambda }{C} \right)\,\left(e^{C\,t} -1 - C\,t - \frac{1}{2}\,C^2\,t^2 \right)\ , $$
where $C$ and $D$ are the integration constants.

In the limit $C\rightarrow0$, we have $v=0$ and thus we are back to the K\"ahler-Einstein metrics, \emph{i.e.} the K\"ahler-Einstein extension of Eguchi-Hanson (the correspondence between the parameters is then $D=B$)\footnote{This shows that the four parameters of the general solution of (\ref{eq}) can not be $A$, $B$, $C$ and $D$ as these are not independent.}.

\section{Two-loop renormalisation}
The two-loop divergences, first computed by Friedan \cite{Fr}, are
$$Div^2_{ij}=-\frac{\hbar^2\,T}{8\,\pi^2\epsilon}R_{is,tu}{R_j}^{s,tu}\ ,\qq d=2-\epsilon\ .$$
In order to re-absorb these divergences, the counter-terms may come from the renormalisation of the coupling $T$ and the fields $\vec\phi$, but also from the renormalisation of the parameters that were let in the metric at one-loop. For example, if one starts with the Einstein metric (\ref{TN}), one should allow for counter-terms renormalising the parameters $M$, $n$. In general, if we define such parameters as $\rho_c$, the theory will be renormalisable at two loops iff one can find some vector ${\tilde v}={\tilde v}(t)$ and some constants $\tilde \lambda$ and $\chi_c$ such that
\begin{equation}\label{RR}
\frac12 R_{is,tu}{R_j}^{s,tu}=\tilde \lambda\,g_{ij}+\chi_c\,\pt_{\rho_c}g_{ij}+D_{(i}{\tilde v}_{j)}\ .
\end{equation}
We will show that, except for the few particular cases where the metric is homogeneous\footnote{It was proven in \cite{BBB} that homogeneous metrics are always renormalisable to all loop order.}, the $SU(2)\times U(1)$ Einstein and K\"ahler metrics do not give in a direct way two-loop renormalisable $\sigma$-models.

\vspace{0.3cm}
\begin{flushleft}
\textbf{Einstein metrics :}
\end{flushleft}

In the vierbein basis, one can compute the two-loop divergences for the metric given in (\ref{TN}) and find~:
$$\frac12 R_{am,np}R_{bm,np}=3\left(\frac{{\left( M - n \right) }^2}{{\left( n - t \right) }^6}  + 
  \frac{{\left( M + n + \frac{8\,n^3\,\lambda }{3} \right) }^2}{{\left( n + t \right) }^6}+ \frac{{\lambda }^2}{9}\right)\delta_{ab}\ .$$
Quite surprisingly, the two-loop divergences are conformal to the original metric.

Relation (\ref{RR}) in the vierbein basis becomes
$$\frac12 R_{am,np}R_{bm,np}=\frac12\,\tilde\lambda\,\delta_{ab}+{E_{a}}^j(\chi_M\,\pt_M+\chi_n\,\pt_n)E_{bj}+\frac12\,D_{a}\tilde v_{b}+(a\leftrightarrow b)\ ,$$
were $E_{ai}$ is defined by $e_a=E_{ai}\,d\phi^i$. As for the one-loop renormalisation conditions (\ref{eq}), this last relation gives us three equations. These can easily be reduced to two  by eliminating $\tilde v$. The remaining equations will only depend on the variable $t$ and on the constants $\tilde \lambda$, $\chi_n$ and $\chi_M$. As these must vanish irrespectively of the values taken by $t$, one can show that they will be verified in only two particular cases where $M$ and $n$ are fixed such that
$$M^2=n^2=-\frac{3}{4\,\lambda} \qq\hbox{or}\qq M=n=0\ .$$
In both cases, (\ref{RR}) will be satisfied with $\dst\tilde\lambda=\frac{\lambda^2}{3}$ and $\chi_M=\chi_n=\tilde v=0$, but it is not surprising as these choice for $M$ and $n$ are the one which enlarge the $SU(2)\times U(1)$ isometries to $SO(5)$, making the metric homogeneous (de Sitter metric).

\vspace{0.3cm}
\begin{flushleft}
\textbf{K\"ahler metrics :}
\end{flushleft}

Proceeding as for the Einstein metrics, one can compute the two-loop divergence using the metric (\ref{ka}). Once again, the parameters $C$ and $D$  must have special values for the action to be two-loop renormalisable. Indeed, one must have $(C=2\,\lambda,D=0)$ or $(C\rightarrow0,D=0)$. In the first case, we recover flat space. In the second case, we get the Fubiny-Study metric on $P_2(\mathbb{C})$ and its non compact partner which are also two-loop renormalisable with $\tilde \lambda=\frac23\,\lambda^2$ and $\tilde v=0$.

\vspace{0.5cm}

The Einstein and K\"ahler metrics with no more isometries than $SU(2)\times U(1)$ are therefore not renormalisable in the minimal scheme at two loops. This could of course be cured by adding some infinite deformation of the metric itself as in D. Friedan's approach to $\sigma$ models quantisation, but it is the author belief that a finite deformation keeping the isometries, as explained in \cite{BC}, would be sufficient\footnote{Here, one should start with the \emph{general metric}, solution of (\ref{eq}), if no new parameters is a required condition for the renormalisation process.}.

\section{The dual metric}

We dualise the initial metric (\ref{meti}) over the $SU(2)$ isometries, keeping aside the $U(1)$. Practically, it consists in dualising the three dimensional metric \cite{He}
$$g_3=t\,({\sigma_1}^2+{\sigma_2}^2)+\gamma(t)\,{\sigma_3}^2\ ,$$
leaving the term $\alpha(t)\,dt^2$ unchanged. If we define the new fields of the dual metric $\lambda^i$, $i\in \{1,2,3\}$, the dual theory of $g_3$ will writes, in light-cone coordinates~:
$$\hat S_3=\frac{1}{T}\int dx^2\,\hat {G_3}_{ij}\,\pt_+\lambda^i\,\pt_-\lambda^j\ ,$$
where
$$\hat {G_3}_{ij}=
{\left(
\begin{array}{ccc}
 t  & \lambda_3 & -\lambda_2 \\[2mm]
-\lambda_3 & t  &  \lambda_1 \\[2mm]
 \lambda_2 & -\lambda_1 & \gamma(t)
\end{array}
\right)^{-1}}_{ij}\ .
$$
After the following change in coordinates :
$$\lambda_1 = y\,\sin(z),\qq
\lambda_2 = y\,\cos(z),\qq
\lambda_3 = r,$$
one has for the total dual metric ${\hat g}=\alpha(t)\,dt^2+\hat {G_3}_{(ij)}\,d\lambda^i\,d\lambda^j$ :
\begin{equation}\label{metf}
{\hat g}=\alpha (t)\,dt^2  + 
  \frac{r^2+t^2}{\Delta}\,\left(dr+\frac{r\,y}{r^2+t^2}\,dy\right)^2+ \frac{t}{r^2+t^2}\,dy^2 + 
  \frac{t\,y^2\,\gamma (t)}
   {\Delta}\,dz^2 
\end{equation}
where
$$\Delta=y^2\,t + \left( r^2 + {t}^2 \right) \,\gamma (t)\ .$$

The torsion is defined by $T=\frac12\,dH$ where $H=\frac12\,\hat {G_3}_{[ij]}\,d\lambda^i\wedge d\lambda^j$ is the torsion potential 2-form~:
\begin{equation}\label{torsion}
H=d\left(z\,dr\right)+\frac{\left(r^2+t^2\right)\,\gamma(t)}{\Delta} \,dr\wedge dz +
  \frac{r\,y\,\gamma (t)}{\Delta}\,dy\wedge dz\ .
\end{equation}

We define ${\hat g}_{ij}$ as the tensor associated to the metric (\ref{metf}) and ${\hat h}_{ij}$ as the torsion potential. Let ${\hat G}_{ij}={\hat g}_{ij}+{\hat h}_{ij}$ and ${\hat Ric}$ be the new Ricci tensor which is not symmetric anymore because of the presence of torsion in the dualised model. Eventually, the dualised action of our $SU(2)\times U(1)$ theory is, in light-cone coordinates ~:
\begin{equation}\label{actionf}
\hat S=\frac{1}{T}\int dx^2\,\hat {G}_{ij}\,\pt_+\hat \phi^i\,\pt_-\hat \phi^j\ ,
\end{equation}
where the coordinates are $\{\hat \phi^0=t,\hat \phi^1=r,\hat \phi^2=y,\hat \phi^3=z\}$. It could be useful to notice that
$$\det \hat g=\frac{t^2\,y^2}{\Delta^2}\,\alpha(t)\,\gamma(t)\ .$$
It was proved in \cite{aal1} that the dualised Eguchi-Hanson model is conformally flat. We have checked that, in the class studied here, this is the \emph{only} case where the Weyl tensor vanishes.

\vspace{0.3cm}
\begin{flushleft}
\textbf{The \mathversion{bold}$SO(3)$\mathversion{normal} dual of Schwarzschild  :}
\end{flushleft}
Among all the $SU(2)\times U(1)$ metrics, the Schwarzschild one has an interesting peculiarity as its dual can be obtained in two ways. Indeed, in the original metric (\ref{schwar}), due to the split of ${\sigma_3}^2$, the $SU(2)$ isometries appear only in the $({\sigma_1}^2+{\sigma_2}^3)$ term. One can therefore first dualise the ``sub-metric'' corresponding to this last term and then add the $dt^2$ and $d\Psi^2$ terms in order to obtain the dualised Schwarzschild metric. Doing this, only two Lagrange multipliers $\lambda^i$ will appear during the dualisation procedure \cite{He}. But it is still possible to obtain it by first dualising the metric (\ref{TN}) and \emph{then} taking the appropriate limit ($n\rightarrow 0$). As $\gamma(t)\rightarrow 0$, one has first to make the change of coordinates $dz=\frac{d\Psi}{2\,n}$ before taking the limit. Doing this, one gets for $\hat g$ :

$$\hat g=\frac{1}{\dst1-\frac{2\,M}{t}-\frac{\lambda}{3}\,t^2}\,dt^2+\frac{r^2+t^4}{t^2\,y^2}\left(dr+\frac{r\,y}{r^2+t^4}\,dy\right)^2
+\frac{t^2}{r^2+t^4}\,dy^2+\left(1-\frac{2\,M}{t}-\frac{\lambda}{3}\,t^2\right)\,d\Psi^2\ .$$
Finally, by making the coordinate change $y=\sqrt{s^2-r^2}$, we get~:
\begin{equation}\label{SM}
\hat g=\frac{1}{\dst1-\frac{2\,M}{t}-\frac{\lambda}{3}\,t^2}\,dt^2+\left(1-\frac{2\,M}{t}-\frac{\lambda}{3}\,t^2\right)\,d\Psi^2
+\frac{1}{t^2\,\left( s^2 -r^2 \right) }\,\left(t^4\,dr^2+s^2\,ds^2\right)\ .
\end{equation}
In the special case $\lambda=0$, we recover the $SO(3)$ dual of Schwarzschild which was one of the first examples for non-abelian duality \cite{oq}. 
While making $n\rightarrow 0$, the torsion potential 2-form $H$ (\ref{torsion}) writes as $d\left(\frac{\Psi\,dr}{2n}\right)+O(n)$, and therefore, as $H$ is only defined up to a total derivative, the torsion vanishes, which is consistent with the result found in \cite{oq}.

\vspace{0.3cm}
We will now address the question of the one loop renormalisability of the dual theory $\hat S$.

\section{One-loop renormalisation of the dual metric}

We want to prove that the one-loop renormalisation property does survive to the dualisation process. In other words, if the torsionless action (\ref{actioni}) is quasi-Einstein, then so is the action (\ref{actionf}). In the presence of torsion, this now means that one can find some constant $\hat \lambda$ and some vectors $\hat v$ and $\hat w$ such that
\begin{equation}\label{QE}
{\hat Ric}_{ij}={\hat \lambda}\,{\hat G}_{ij}+D_j\hat v_i+\pt_{[i}{\hat w}_{j]}\ .
\end{equation}
This equality gives a system of equations much more complicated than (\ref{eq}), but what is important is that now $\alpha(t)$ and $\gamma(t)$ are not considered as unknown functions. Furthermore, as we suppose the original metric to be quasi-Einstein, the system (\ref{eq}) is assumed to be verified and one can easily derive from it, in an algebraic way, the three functions $A$, $B$ and $C$ such that~:

\begin{equation}\label{ABC}
\left\{
\begin{array}{rll}
\dst \alpha'(t)  &=& \dst A\left(t,\alpha(t),\gamma(t),v_0(t),v_0'(t)\right)\ ,\\
\dst \gamma'(t)  &=& \dst B\left(t,\alpha(t),\gamma(t),v_0(t),v_0'(t)\right)\ ,\\
\dst \gamma''(t) &=& \dst C\left(t,\alpha(t),\gamma(t),v_0(t),v_0'(t)\right)\ .
\end{array}
\right.
\end{equation}
The procedure is the following : we choose some ansatz for $\hat \lambda$, $\hat v$ and $\hat w$ and express relation (\ref{QE}). Then, in this last expression, we replace each occurrence of $\alpha'(t)$, $\gamma'(t)$ and $\gamma''(t)$ by its expression in (\ref{ABC}) and check if (\ref{QE}) holds.

We have checked that (\ref{QE}) is verified taking
\begin{equation}\label{choix}
\left\{
\begin{array}{rll}
\dst {\hat \lambda} &=& \dst \lambda \ ,\\
\dst {\hat v}_i     &=& \dst -2\,\lambda\,{\hat g}_{ij}\,X^j + D_i\log\Delta +v_i\ ,
\\
\dst {\hat w}_j      &=& \dst -2\,\lambda\,X^j\,{\hat G}_{ji}\ ,
\end{array}
\right.
\end{equation}
where $X$ is defined by $X=r\,\pt_r+y\,\pt_y$.

Conversely, let us now suppose that $\hat \lambda$, $\hat v$ and $\hat w$ are defined by (\ref{choix}) where $\lambda$ and $v$ are supposed to be arbitrary. It is possible to show that if (\ref{QE}) holds, then the original metric is quasi-Einstein with $Ric_{ij}=\lambda\,g_{ij}+D_{(i}v_{j)}$. In order to demonstate this, we first define the three functions $f_A(t)$, $f_B(t)$ and $f_C(t)$ such that :
\begin{equation}\label{ABC2}
\left\{
\begin{array}{rll}
\dst \alpha'(t)  &=& \dst A\left(t,\alpha(t),\gamma(t),v_0(t),v_0'(t)\right)+f_A(t)\ ,\\
\dst \gamma'(t)  &=& \dst B\left(t,\alpha(t),\gamma(t),v_0(t),v_0'(t)\right)+f_B(t)\ ,\\
\dst \gamma''(t) &=& \dst C\left(t,\alpha(t),\gamma(t),v_0(t),v_0'(t)\right)+f_C(t)\ .
\end{array}
\right.
\end{equation}
Assuming that (\ref{QE}) holds, and after having replaced each occurence of $\alpha'(t)$, $\gamma'(t)$ and $\gamma''(t)$ by its value in (\ref{ABC2}), we get some equation system where the unknowns are the functions $f_X(t)$. As this last system must hold irrespectively of the values taken by $r$ and $y$ which are free variables, one can then prove that $f_A(t)=f_B(t)=f_C(t)=0$. This shows that (\ref{ABC}) holds and therefore the quasi-Einstein property of the original metric.

We have proven, for arbitrary functions $\alpha(t)$ and $\gamma(t)$, the equivalence
\begin{equation}\label{equiv}
Ric_{ij}=\lambda\,g_{ij}+D_{(i}v_{j)}\qq\Longleftrightarrow\qq{\hat Ric}_{ij}={\hat \lambda}\,{\hat G}_{ij}+D_j\hat v_i+\pt_{[i}{\hat w}_{j]}\ ,
\end{equation}
where $\lambda$, $\hat \lambda$, $v$ and $\hat v$ are related by (\ref{choix}).
\pagebreak
\begin{flushleft}
\textbf{Remarks :}
\end{flushleft}
\begin{itemize}
\item The cosmological constant does not change through the dualisation process as it was already proved for T-dualised homogeneous metrics \cite{CV}. That means that the coupling will renormalise in exactly the same way that in the initial theory : the one-loop Callan-Symanzik $\beta$ function is the same for the initial and dualised $SU(2)\times U(1)$ theories.

\item As one could expect, the coordinate $t$ which was a spectator coordinate during the dualisation process plays a special role : ${\hat w}_t=0$ and, up to the $D_t\log\Delta$ term, ${\hat v}_t$ and $v_t$ are equal.

\item The $SU(2)$ symmetries where lost during the dualisation process, so at the end, there is just a $U(1)$ symmetry left and therefore the Killing  $\pt_z$ is unique. Indeed, $\hat v$ and $\hat w$ are defined up to this Killing vector, which dual 1-form is $K=\frac{y^2\,t\,\gamma (t)}{\Delta}\,dz$. One then has $D_{(i}K_{j)}=0$ and $D_{[j}K_{i]}+\pt_{[i}K^s{\hat G}_{sj]}=0$.

\item  One can adress the question of the unicity of $\hat \lambda$, $\hat v$ and $\hat w$ which satisfy (\ref{QE}). There will be multiple solutions if one can find some $\Lambda$, $V$ and $W$ such that 
$$\Lambda\,{\hat G}_{ij}+D_jV_i+\pt_{[i}W_{j]}=0\ .$$
On the one hand, $\hat w$ alone is obviously defined up to a gradient while $\hat v$ and $\hat w$ together are defined up to the Killing vector $K$ ; on the other hand, equivalence (\ref{equiv}) shows that if multiple solutions exist for $\hat \lambda$ and $\hat v$ in the dualised metric, then such ambiguity will appear for the original metric. We have checked that, in our case of $SU(2)\times U(1)$ metrics, only flat metric leads to such possibilities\footnote{For flat space ($\beta(t)=\gamma(t)=1/\alpha(t)=t$), we have $\lambda\,g_{ij}+D_{(i}v_{j)}=0$ with $v=-2\,\lambda\,dt,\ \forall\lambda\in\mathbb{R}$.} . Therefore, except for this trivial original metric and up to the already noticed freedom in $\hat v$ and $\hat w$, (\ref{choix}) is the unique solution of (\ref{QE}).

\item The $SO(3)$ dual of the Schwarzschild metric (\ref{SM}) gives us a nice example of a torsionless quasi-Einstein metric with a $U(1)$ as minimal isometry.
\end{itemize}

\section{Conservation of the K\"ahler property}

Bakas and Sfetsos decribed, for SUSY applications, how the complex structures were changed
when hyper-K\"ahler metrics were T-dualised \cite{basf}. We propose here to show that when one starts with 
the original metric (\ref{ka}), the dual partner is still K\"ahler.

If we define
$$\hat \si_i=-\hat G_{si}\,d\hat\phi^s,$$
it is possible to write the dual metric of (\ref{ka}) under the specific shape
$$\hat g=\frac1{\gamma(t)}\,dt^2+t\,(\hat{\si_1}^2+\hat{\si_2}^2)+\gamma(t)\,\hat{\si_3}^2~.$$
One can then check that the 2-form
$$\hat\rho=dt\wedge \hat\si_3+t\,\hat\si_1\wedge\hat\si_2=\frac12\hat {\cal J}_{ij}\,d\hat\phi^i\wedge
d\hat\phi^j $$
is a K\"ahler form {\bf with torsion} for the dual metric. Indeed, for the almost complex structure
$\hat {\cal J}$, we have 
$$\left\{
\begin{array}{l}
\hat {\cal J}_{is}\,\hat {\cal J}^{sj}=-{\delta_i}^j~,\\[2mm]
\hat {\cal J}_{(ij)}=0~,\\[2mm]
D_i\hat {\cal J}_{jk}=0~,
\end{array}\right.$$
where $D$ is the covariant derivative with torsion.
One should notice here that, in the presence of torsion, the closing condition on the K\"ahler form is replaced by
$$d\hat\rho=(\star\, dH)\wedge \hat\rho~.$$
The torsion potential 2-form $H$ is given by the equation (\ref{torsion}).

\section{Concluding remarks}

We have considered all of the four dimensional non homogeneous metrics with an isometry group $SU(2)\times U(1)$. We have shown that the dual partners are quasi-Einstein (with torsion) iff the original metrics are quasi-Einstein (without torsion). Let us emphasize that this was possible despite the fact that the explicit form of these metrics are not all known yet. 

In \cite{kt}, it was proven that, in the minimal dimensional scheme, the dualised $SU(2)$ principal $\sigma-$model is not two-loop renormalisable although this property holds for its original model. Here, the one-loop renormalisability remains although the starting models are not in general two-loop renormalisable. This is another suggestion that the renormalisability beyond one loop for the original and dualised models are not linked. Indeed, it is our ansatz that for the dualised models investigated here, one could still define a proper theory up to two loops. This could be achieved by adding some \emph{finite} deformation to the dualised metric, as it was done in \cite{BC} for the $SU(2)$ principal $\sigma-$model, irrespectively of the two-loop renormalisability of the original theory.

\vspace{3mm}

\noindent{\bf Acknowledgments :} I am indebted to G. Valent for suggesting this work and to G. Bonneau for enlightening discussions and remarks.

\end{document}